\documentclass{article}    

\usepackage{graphicx}
\newcommand{\bra}[2]{\ensuremath{\langle#1|_{#2}}}
\newcommand{\ket}[2]{\ensuremath{|#1\rangle_{#2}}}

\title{\textbf{Quantum Consciousness}}  
\author{Richard Mould\footnote{Department of Physics and Astronomy, State University of New York, Stony Brook,
\mbox{New York} 11794-3800}}  
\date{State University of New York, Stony Brook, N.Y.}    
   
\begin{document}             

\maketitle              

\begin{abstract}

In a previous paper, the author proposed a quantum mechanical interaction that would insure that the
evolution of subjective states would parallel the evolution of biological states, as required by von NeumannÕs
theory of measurement.  The particular model for this interaction suggested an experiment that the author has now
performed with negative results.   A modified model is outlined in this paper that preserves the desirable
features of the original model, and is consistent with the experimental results.  This model will be more
difficult to verify.  However, some strategies are suggested.  

\vspace{4mm}


\end{abstract}

\section{Introduction}
	
	For conscious states and brain states to mirror one another in any species, thereby establishing what von
Neumann calls a psycho-physical parallelism, these intrinsically different states must evolve together and
interact with one other during their time of evolution.  Standard physics makes no provision for an interaction
of this kind, but a quantum mechanical opening for an objective/subjective interaction is shown to exist, and is
described in previous papers.\cite{M95}  \cite{M98}

	Our theory of subjective evolution calls for the existence of a Central Mechanism ($CM$) within an evolving
organism, which contains presently unknown components of the nervous system.  The function of a $CM$ is to reduce
quantum mechanical superpositions within the nervous system, and to simultaneously give rise to a conscious
experience of the eigenvalues of the reduction.  This accords with von NeumannÕs requirement that a quantum
mechanical state reduction is accompanied by an observer's conscious experience of the measured variables.  At the
present time, no one knows what there is about a conscious organism that gives rise to either consciousness or
state reduction. We simply combined these two mysteries inside the $CM$, thereby placing our ignorance in a
\mbox{black-box} so we can ask another question, namely: how do physical and mental states evolve interactively
to insure the psycho-physical parallelism?

The model in references 2 and 3 requires that a conscious organism spontaneously creates a profusion of
macroscopic quantum mechanical superpositions consisting of different neurological configurations.  A mechanism for
this generation is \mbox{proposed} by H. Stapp.\cite{HS}  The result is a superposition of different neurological
states, each of which may be accompanied by a different subjective experience.  A reduction to a single eigenstate
is not assumed to be triggered microscopically along the lines of Ghirardi-Rimini-Weber\cite{GRW}; but rather, it
is assumed to occur in response to a macroscopic event.  It occurs the moment an emerging subjective state becomes
actively conscious in one of the macroscopic neurological components of a Stapp superposition.  The consciousness
that is associated with such a reduction is assumed to fade the moment reduction is complete, and the
resulting subjective \emph{pulse} is supposedly followed by similar pulses in rapid succession.  This can make the
subject aware of an apparent continuum of consciousness.   

Presumably, any reduction of this kind is accompanied by a reduction of all other parts of the organism as well as
all those parts of the external world that are correlated with it.  This means that a second observer, coming on
the heels of the first, will make an observation in agreement with the first.  More formally, a measurement
interaction establishes correlations between the eigenstates \ket{a_i}{} of some apparatus (with discrete variables
$a_i$), eigenstates of a first observer\ket{\Phi_i}{}, and eigenstates of a second observer \ket{\Theta_i}{},
such that the total state prior to reduction is given by
$\ket{\Psi}{}=\Sigma_iC_i\ket{a_i}{}\ket{\Phi_i}{}\ket{\Theta_i}{}.$ The coefficient $C_i$ is the probability
amplitude that the apparatus is in state \ket{a_i}{}.  Let the first observer become consciously aware of the
apparatus variable $a_k$.  The resulting reduction is a projection in Hilbert space that is found by applying the
projection operator of that observer \ket{\Phi_k}{}\bra{\Phi_k}{} to the total state.

\begin{tabbing}
 xxxxxxxxxxxxxxxxxx\=xxxxxxxxxxxxxxxxxxxxxxxxxxxxxxxxxxxx\=          \kill                
 \>$\ket{\Phi_k}{}\bra{\Phi_k}{}\ket{\Psi}{}=C_k\ket{a_k}{}\ket{\Phi_k}{}\ket{\Theta_k}{}$		\>(1st reduction) 
\end{tabbing}
 Let the second observer then become consciously aware of the apparatus variable $a_m$.  The subsequent reduction
is found by applying the projection operator of that observer \ket{\Phi_m}{}\bra{\Phi_m}{} to the first reduction.

\begin{tabbing}
 xxxxxxxxxxx\=xxxxxxxxxxxxxxxxxxxxxxxxxxxxxxxxxxxxxxxxxxx\=          \kill                

\>\ket{\Theta_m}{}\bra{\Theta_m}{}$C_k$\ket{a_k}{}\ket{\Phi_k}{}\ket{\Theta_k}{}=$\delta_{km}C_k$\ket{a_k}{}\ket{\Phi_k}{}\ket{\Theta_m}{} 
\>(2nd reduction) 
\end{tabbing}
Only if $m=k$ is the probability non-zero that the second observer will make a measurement.  The second observer therefore
confirms the results of the first observer that the apparatus has been left in the eigenstate \ket{a_k}{}.

	Again, many of the particulars of a reduction (such as its nonlinearly) are ignored in this paper so we can
concentrate on the influence of subjective states on physiological states.  To this end we require  that
\emph{when the emerging subjective states of a  neurological superposition are different from one another, they
will generally exert an \mbox{influence} on their relative probability amplitudes that is a function of that
difference.}  In particular, we imagine that when a ÔpainfulÕ subjective state emerges in
\mbox{superposition} with a ÔpleasurableÕ subjective state, the probability amplitude of the painful state will be
decreased relative to the probability amplitude of the \mbox{pleasurable} state.  

No currently known observation contradicts this conjecture, for no \mbox{previously} reported experiment deals
specifically with the creation of different observers \mbox{experiencing} different degrees of pain, arising on
different components of a \mbox{quantum} mechanical superposition.    

Let $N$ in fig.\ 1 represent the nervous system of the first primitive organism that makes a successful use of the
subjective experience of ÔpainÕ.  In a previous paper we imagine this creature to be a fish. It is
supposed that the fish makes contact with an electric probe, at which time its nervous system splits into a
superposition ($via$ the Stapp mechanism) consisting of a withdrawal behavior $W$ that is accompanied by [no
pain], and a continued contact behavior $C$ that is accompanied by [pain].  The probability of survival of each
component in this highly artificial model is initially assumed to be 0.5.  However, because of the
hypothetical influence of subjective pain on probability amplitudes, only the withdrawal state is assumed
to survive the reduction in this idealized example.  State reduction in fig.\ 1 is represented by the
horizontal arrow.  If $W$ is further-more a good survival strategy from the point of view of evolution, then
the association $W$[no pain] and $C$[pain] will serve the species well, whereas a wrong association $W$[pain] and
$C$[no pain] will lead to its demise.

\begin{figure}[t]
\centering
\includegraphics[scale=0.8]{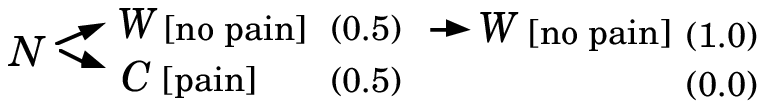}
\center{Figure 1}
\end{figure}

It does not matter to the above argument if the variables are Ôpleasure/painÕ or some other range of subjective
experiences.  If a subjective experience like `A' increases the probability amplitude of an escape
behavior, and if a \mbox{subjective} experience like `B' diminishes the probability amplitude of that
behavior, and if the escape is one that moves the creature away from something that is dangerous to its health,
then a distant descendent  will experience `A' associated with life supporting escapes, and `B' associated
with life threatening failures-to-escape.  It is apparent that the quality of the experience does not matter.  We
require only that the subjective experience in question has a predictable plus or minus effect on the probability
amplitudes within a superposition, and the survival mechanisms of evolution will do the rest.  They will insure
that the eventual subjective life of a surviving species mirrors its experiences in a definite and predictable
way Ð thereby establishing a reliable psycho-physical parallelism.  

\begin{figure}[h]
\centering
\includegraphics[scale=0.8]{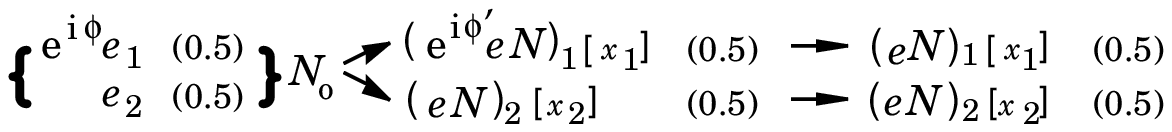}
\center{Figure 2}
\end{figure}

  We assume that ordinary $perception$ do not have this effect.  They do not give rise to the hypothetical
feedback.  In fig.\ 2 we imagine the existence of an externally imposed two component superposition consisting of
environments $e_{1}$ and $e_{2}$, which is produced by using, say, a $\beta$ source.  The two environments
are assumed to have equal probability, and are allowed to interact with the subjectÕs nervous system given by
$N_{0}$.  Before a reduction can occur, two conscious states emerge from the interaction represented by the
superposition of $(eN)_{1}[x_{1}]$ and $(eN)_{2}[x_{2}]$, where the conscious part shown in brackets is the
observed \mbox{eigenvalue} $x$ associated with components 1 and 2.  Since we require that an observer of the
ÔperceivedÕ variable $x$ cannot affect the probability of $x$, the pure state reduces to a
mixture having the same probability as the initial superposition (horizontal arrows in fig.\ 2).  State $e_i$
represents the relevant laboratory apparatus together with the wider environment with which it is entangled.  The
phase angles $\phi$ and $\phi'$ are definite, but they are not localized to manageable parts of the
apparatus.\cite{JZ}.  We call them "arbitrary" in this paper to indicate that their values are not practically
calculable, and to emphasize the lack of coherence between these "macroscopic" components.

\begin{figure}[t]
\centering
\includegraphics[scale=0.8]{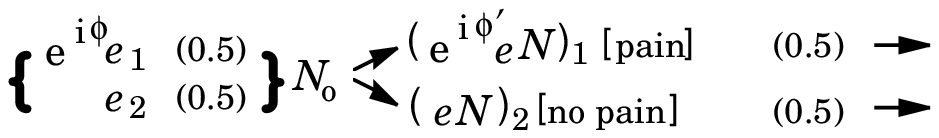}
\center{Figure 3}
\end{figure}  

On the other hand, if ÔpainÕ were the variable in fig.\ 2 rather than the \mbox{externally} perceived variable $x$,
it is suggested by our hypothesis that the resulting mixture might no longer be a 50 - 50 split.  This possibility
is represented in fig.\ 3, where the final mixture probabilities are left unspecified because they must be
discovered by observation.  The author has now performed an experiment of this kind with the result reported
below.       

\section{The Experiment}

Two scalers L and R recording local background radiation are placed \mbox{side-by-side} in fig.\ 4.  Their outputs
are fed to a selector box that chooses channels L or R, depending on which is the first to record a single count
after the selector has been turned on.  A 20 V signal is then emitted from the output of the chosen channel.  The
output on the R-channel is unused, but the L-output closes a relay that puts 80 volts across two metal bars.  Two
seconds after the selection, an L or R-light goes on indicating which channel was selected.  A finger placed
across the metal bars will receive a painful 80V shock when the L-channel is selected. 

This apparatus allows us to carry out the experiments diagramed in figs. 2 and 3.  If the selector is initiated
in the absence of an observer, we say that the system will become a macroscopic superposition given by
$(e^{i\phi}e_{1} + e_{2})$, where $e_{1}$ is the entire apparatus following an L-channel activation, and $e_{2}$
is the entire apparatus following an R-channel activation.  The incoherence of the two components
(represented by the arbitrary angle $\phi$) is generally understood to mean that the system is indistinguishable
from a classical mixture, since interference between these macroscopic components is not possible.  
However, for \mbox{reasons} given in previous papers, we claim that the final state is really an \mbox{incoherent}
quantum mechanical superposition rather than a classical mixture.\footnote{The uncertainty associated with a
classical mixture state represents an outsider's ignorance, whereas a pure quantum mechanical state superposition
represents an uncertainty that is intrinsic to the system (see ref. 1, pp. 1622, 1624; and ref. 2, bottom of p.
1703).  Following von Neumann, we assume that the initial intrinsic uncertainty (concerning which of the scalers
fires first) will remain an intrinsic uncertainty until it is reduced by ``observation".  Hence, the apparatus will
remain a macroscopic pure state quantum mechanical superposition until an observation occurs.}  The lack of
interference between the components has no bearing on our result because the hypothetical effect described in this
paper relates to, and directly affects,  probability amplitudes only.  The effect we are looking for should be
observable with or without coherence between L and R.  

\begin{figure}[t]
\centering
\includegraphics[scale=0.8]{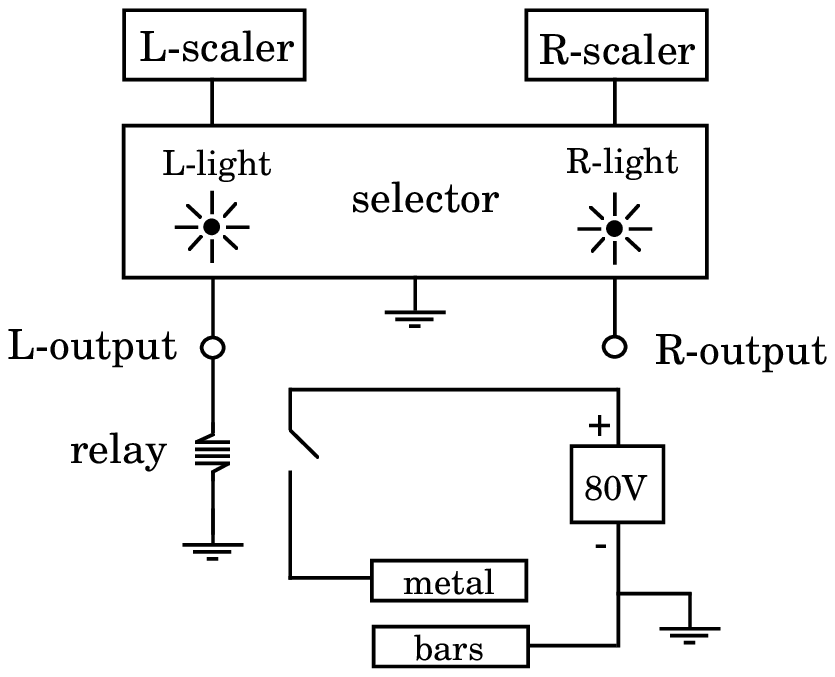}
\center{Figure 4}
\end{figure} 

If an observer is present and exposed only to the L-light or the R-light, then a reduction will occur like the
one in fig.\ 2, where eigenvalues $x_{1}$ and $x_{2}$ represent a conscious experience of one or the other of those
lights.  If the observer is exposed only to a conscious experience of ``pain or no pain" through his finger across
the metal bars, then a reduction like the one in fig.\ 3 will occur.  This experiment may not appear to be quantum
mechanical, but it is quantum mechanical by virtue of the particular hypothesis that is being tested in fig.\ 3.

The equipment in fig.\ 4 was used for a total of 2500 trials, each consisting of two parts.  The authorÕs finger
was first placed across the metal bars, the selector was turned on, and a ÔshockÕ or Ôno shockÕ was recorded
before the lights were observed.  In the second part of each trial the finger was replaced by an equivalent
resistance, the selector was again initiated, and the appearance of the L or R channel light was recorded.  

\vspace{8pt}

Total number of trials . . . . . . . . . . . . . . . . . . . . . . . . . .  $N$ = 2500

Number of shocks received in the first part . . . . . . . . . . . . . $N_{S}$ = 1244

Number of times the L-light went on in the second part . . . . . $N_{L}$ = 1261 

\vspace{8pt}

\noindent
There are three possible outcomes of a single trial.   Either the difference \mbox{$N_{L}$ - $N_{S}$} increases, or
it decreases, or it remains the same.  The three possibilities are represented by the variables $u$ (increase)
occurring with a probability $p$, and
$d$(decrease) with a probability $q$, and $e$ (remain the same) with a probability $r$.  It was found in the
experiment that $u$ = 632 and  $d$ = 615 after 2500 trials.  

If we approximate $p_{0} = N_{L}/N$ to be the probability that the left channel fires in the second part of each
trial (absent the finger), and $q_{0} = 1 - p_{0}$  to be the probability that the right channel fires in the
second part of each trial, then 

\begin{tabbing}
 xxxxxxxxxxxxx\=xxxxxxxxxxxxxxxxxxxxxxxxxxxx\=          \kill                
 \>$p_{0} = 1261/2500 = 0.5044$		\>$q_{0} = 0.4956$ 
\end{tabbing}

\noindent
Assuming as a null hypothesis that there is no statistical difference between the displacement of a
finger across the metal bars and an equivalent resistor, we have $p = p_{0}q_{0}$, $q = q_{0}p_{0}$, and $r =
p_{0}^{2} + q_{0}^{2}$,  
giving

\begin{tabbing}
xxxxxxxxxxxxx\=xxxxxxxxxxx\=xxxxxxxxxxx\=\kill
\>$p$ = 0.2500 \>$q$ = 0.2500 \>$r$ = 0.5000
\end{tabbing}
The variances of $(u + d)$ and $(u - d)$ are

\begin{eqnarray}\nonumber
\sigma^{2}(u + d) =\ <(u + d)^{2}> - <u + d>^{2} = \sigma^{2}(u) + \sigma^{2}(d) + X\\  
\sigma^{2}(u - d) =\ <(u -d)^{2}> - <u - d>^{2} = \sigma^{2}(u) + \sigma^{2}(d) - X    \nonumber
\end{eqnarray}
therefore
\begin{eqnarray}\nonumber
\sigma^{2}(u - d) = 2\sigma^{2}(u) + 2\sigma^{2}(d) - \sigma^{2}(u + d)\\
= 2p(q + r)N + 2q(p + r)N - r(p + q)N	    \nonumber
\end{eqnarray}
or
\begin{eqnarray}\nonumber
\sigma(u - d) = [[4pq + r(p + q)]N]^{1/2} = [N/2]^{1/2} = 35.4
\end{eqnarray}

Our alternative hypothesis is that $u - d$ is significantly different from 0.  But from the data, $u-d =
N_{L}-N_{S} = 17$ after 2500 trials, and this is well within the above the standard deviation around 0.  The
separate variables $u$ and $d$ are also within the standard deviation $\sigma(u) = \sigma(d) = [p(q + r)N]^{1/2} =
21.7$ of their expected value of 625.  

One can always argue that the statistics are inadequate to reveal a significant difference between $u$ and $d$. 
However, they are sufficient to convince the author that the presence of pain on one component of this externally
imposed superposition has no significant effect on the outcome.  We therefore conclude that the reduction in fig.\
3 is not affected by the subjective content of the square brackets in that figure.  

Further details about this experiment can be found in the document "QC Experiment" on the authorÕs home
page.\cite{Mhp}  

\section{Model Modification}

This result forces us to make a distinction between externally imposed superpositions and the superpositions
created internally by a $CM$. 

In fig.\ 5 we let a $CM$ interact with an environmental superposition, where the latter includes all that is not
contained in the $CM$ (including other possible $CM$s within the organism).  The first pair of diverging arrows in
that figure carries the initial product into components $(eCM)_{1}$ and $(eCM)_{2}$.  This reaction goes according
to Schr\"{o}šdinger.  The subsequent pairs of diverging arrows shown in fig.\ 5 denote the appearance of new
superpositions that are produced by the (Schr\"{o}šdinger) process proposed by Stapp.  It is at this point that we
engage the (non-Schr\"{o}šdinger) hypothesis of sect.\ 1 that allows the emerging `subjective' states on different
components of these $CM$-superpositions to vary their own probability amplitudes.  However, we add the further
stipulation that \emph{any such variation can only occur relative to the other components of the originating
$CM$-superposition.}  This restricts the range of states over which the relative amplitude variation can take
place.    

\begin{figure}[h]
\centering
\includegraphics[scale=0.8]{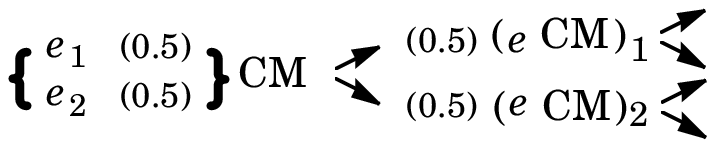}
\center{Figure 5}
\end{figure}

It is thereby required that any variation (due to the hypothetical subjective influence) that takes place within
$(eCM)_{1}$ in fig 5 has no effect on a variation within $(eCM)_{2}$, and vice versa.  More generally, the
probability amplitude of any $CM$ as a whole is not affected by a variation that takes place outside of itself, so
the normalization of each component $(eCM)_{i}$ of the  superposition is preserved.  

This modification is consistent with the results of the experiment in sect.\ 2.  The superposition in our
experiment is represented by the first pair of diverging arrows.  Since it is external and not created by the
\emph{CM}, it cannot, according to the above stipulation, be affected by the subjective content of components 1 and
2.  On the other hand, the second pairs of diverging arrows represent superpositions that \emph{are} created by the
\emph{CM}s 1 and 2,  so they are subject to our hypothesis of sect.\ 1.  On this modified model, subjective
evolution does not rely in any way on externally created superpositions, thereby insuring that the observer of
\mbox{(external)} quantum mechanical systems will always record eigenvalues with the probability predicted by
standard theory.\cite{AS}  At the same time, our hypothetical subjective influence can still be realized within
either one of the $CM$-superpositions in the figure.  

With this modification, the hypothesis becomes much more difficult to demonstrate experimentally.  Although
verification remains possible in principle, it will require a more detailed understanding of the workings of the
nervous system and/or other parts of the body.  

\section{Bio-Active Peptides}

	Neurological communication depends on the diffusion of chemical neurotransmitters across the synaptic junction
between neurons.  There is another communication system within the body that makes use of chemicals that are
produced at one site and received at another; but in this case, the distances between a production and receiver
sites are macroscopic.  About 95\% of these chemical communicators are peptides, which are mini-proteins consisting
of up to 100 amino acids having a maximum atomic mass of   10,000 u.  Their classical \mbox{dimensions} are 
$\Delta$x =10 nm at most, which we assume approximates their size close to the production site.\cite{CP}  
Therefore, Heisenberg tells us that the minimum quantum mechanical uncertainty in the velocity of one of these
free peptides is $\Delta$v =0.63 mm/s.  Peptides are carried through intercellular space by blood and
cerebrospinal fluid.  They do not move very far in a tenth of a second, but in that time the Heisenberg
uncertainty in position of a peptide will be at least $\Delta$s~=~$\Delta$v$\Delta$t = 63 mm.  This is an enormous
uncertainty of position relative to one of the peptide receptor sites which has a size similar to that of the
peptide, and which is often separated from its neighbors by comparable distances.  Therefore, quantum
mechanical uncertainty is an important factor in determining the probability that a given peptide is captured by a
given receptor. 
 
	StappÕs mechanism for introducing quantum mechanical superpositions into the brain relies on the uncertainty in
the position of calcium ions in neuron synapses.  We suggest that peptides represent another possible source
of super-positions that may be just as widespread.  And because peptides play an important role in the
chemistry of the body, they too may have a significant quantum mechanical influence on behavior.  

As with the Stapp mechanism, one might object that the uncertainty associated with the peptideÕs classical
diffusion during its migration will overwhelm the quantum mechanical uncertainty, or that a large number of
migrating molecules will obscure all quantum mechanical effects.  However, the classical uncertainty associated
with many-particle ensembles has only to do with our ignorance of initial conditions.  In reality, the only
uncertainties a receptor will see are those associated with an incoherent quantum mechanical superposition of pure
\mbox{peptide} states.  This superposition will have as many  components as there are peptide molecules involved. 
And since our hypothetical influence acts through the amplitude of these components, the presence of a large
number of independent particles will only increase the hypothetical influence.      

\section{Drugs}

	There are many drugs that can be introduced into the body that will \mbox{compete} with endogenous peptides to
occupy the bodyÕs receptor sites, and some of these drug molecules are small enough to have a very large quantum
\mbox{mechanical} \mbox{uncertainty} of position.  For this reason, peptide/drug superpositions are more promising
for the purpose of experimental manipulation than calcium ion super-positions.  

	For example, $endorphins$ are peptides that unite with special receptors to eliminate pain and/or produce
euphoria.  They and their receptors can be found everywhere in the body, but they are most intensely located in
the limbic \mbox{system} of the brain.  There is a drug called $naloxone$ that is a strong \mbox{competitor} with
the endorphins to occupy the same receptors, and it has the property that it reverses the analgesic/pleasurable
effects of the endorphins.\cite{CP}\cite{SS}\cite{CL}  If \mbox{endorphin} molecules and externally administered
naloxone molecules are in quantum mechanical superposition with one another as their sizes and likely time together
suggests, and if they both compete with one another for successful attachment to the same receptor site, then the
ratio of endorphin attachments to naloxone attachments would (according to our hypothesis) be a function of the
competing subjective states.  Since the difference in subjective effects \mbox{between} these two molecules is
considerable along the pleasure/pain spectrum, an \mbox{experimental} design involving endorphin/naloxone
superpositions appears to offer an
\mbox{opportunity} to test the modified model proposed in sect.\ 3.  The author is not able to propose a specific
experiment at this time, but an approach along these lines seems promising.  

\section{Evolutionary Advantage}

	It was pointed out in a previous paper that our evolutionary mechanism of objective-subjective
interaction (represented by fig.\ 1) does not insure that a creature evolving under its influence will evolve more
quickly or be more successful than a creature evolving strictly as an automaton.  That will be true as well of the
modified model in sects.\ 3-5.  However, it is not unreasonable to suppose that both
conscious evolution and autonomic evolution might work separately and in tandem with one another.  The kinds of
neurological changes that are necessary for autonomic evolution might very well be independent of the kinds of
neurological changes that are necessary for quantum/consciousness evolution.  If that is so, and if these two
processes work in tandem, then the evolution of the organism will be faster than either the autonomic route by
itself, or the conscious route by itself.  One would then be able to say that the introduction of consciousness as
proposed in this paper will always work to the advantage of the organism.

\section*{Acknowledgement} 

The author would like to thank Erlend Graf for his many valuable comments and suggestions.

\end{document}